\documentclass[]{spie}
\usepackage[]{graphicx,amsmath,amsfonts,float}
\usepackage[linesnumbered,ruled, noline]{algorithm2e}

\usepackage[]{fancyhdr,lineno,lastpage}
\newcommand{\hl}{}

\title{Accelerating computed tomographic imaging spectrometer reconstruction using a parallel algorithm exploiting spatial shift-invariance} 
\author{Larz White\supit{a}, W. Bryan Bell\supit{a}, Ryan Haygood\supit{a}
\skiplinehalf
\supit{a}Lockheed Martin Aeronautics, Fort Worth, TX 76108}
\authorinfo{Send correspondence to larz.g.white@lmco.com}

\begin{document} 
\maketitle

\begin{abstract}
Computed Tomographic Imaging Spectrometers (CTIS) capture hyperspectral images in realtime. However, post processing the imagery can require enormous computational resources; thus, limiting its application to non-realtime scenarios. To overcome these challenges we developed a highly parallelizable algorithm that exploits spatial shift-invariance. To demonstrate the versatility of our new algorithm, we developed implementations on \hl{a} desktop and \hl{an} embedded graphics processing \hl{unit} (GPU). To our knowledge, our results show the fastest image reconstruction times reported.
\end{abstract}


\section{Introduction}
Hyperspectral imaging has widespread uses. From commercial applications such as mining, agriculture, and geology\cite{horton2010novel,descour1995computed,vane1988terrestrial,sabatke2002snapshot,Azzam1995,chipman1995,Henderson1995,Horton24,scholl2003phase,luo2007fast} to defense applications, where \hl{it has} been used for Intelligence Surveillance and Reconnaissance (ISR)\cite{horton2010novel,descour1997demonstration,goetz1985imaging,johnson2007snapshot,sabatke2002snapshot,Azzam1995,chipman1995,Henderson1995,vo2014biomedical,descour1995computed,murguia2000compact,wilson2005reconstruction,scholl2003phase}.

The goal of hyperspectial imaging is to obtain \hl{a datacube which is composed of images (the spatial component) of a scene at multiple wavelengths (the spectral component).} For example, \hl{a datacube could be obtained} by taking a picture with a filter placed over the lens, then repeating the \hl{process} with different filters. However, this is a time consuming task that \hl{does not} scale well for dynamic scenes. To overcome these challenges snapshot methods \hl{were} developed \hl{to} encode spatial and spectral information directly into the sensed image. A reconstruction algorithm is then used to \hl{form} the hyperspectral datacube. One instrument for snapshot hyperspectral imaging is CTIS which is robust, simple, fast, and has military successes in missile defense, ISR and target/threat characterization/detection. \cite{horton2010novel, wilson2005reconstruction,scholl2003phase,descour1995computed,scholl2006evaluations}

Even though CTIS captures raw images in milliseconds, \cite{horton2010novel,hagen2007fourier,descour1997demonstration,johnson2007snapshot,hege2004hyperspectral,descour1994non,volin1998high} post processing the imagery requires enormous computational resources, typically taking minutes to hours \cite{thompson2009accelerated,hagen2007fourier}. Since potential applications for CTIS are embedded realtime or near-realtime systems, a better reconstruction algorithm is needed. Previous attempts at speeding up the reconstruction algorithm have been performed by Hagen \cite{hagen2007fourier} and Vose-Horton\cite{horton2010novel,vose2007heuristic} where they developed fast reconstruction algorithms exploiting spatial shift-invariance (hereafter referred to as shift-invariance). \hl{Additionally}, hardware accelerations were performed by Thompson \cite{thompson2009accelerated} and Sethaphong \cite{sethaphong2007large} using cell processors and supercomputers. Our approach is a combination where we developed a highly parallelizable algorithm that exploits shift-invariance and uses a GPU for hardware acceleration. GPUs are common in \hl{modern computers and} provide hardware acceleration while still maintaining \hl{small form factors}. To our knowledge researchers have not utilized GPUs for CTIS reconstruction, but a few hinted at the idea \cite{horton2010novel,thompson2009accelerated,sethaphong2007large}. 

\section{CTIS overview}
\label{sec:ctisoverview}
\begin{figure}[H]
	\begin{center}
    	\begin{tabular}{c}
   		\includegraphics[height=6.5cm]{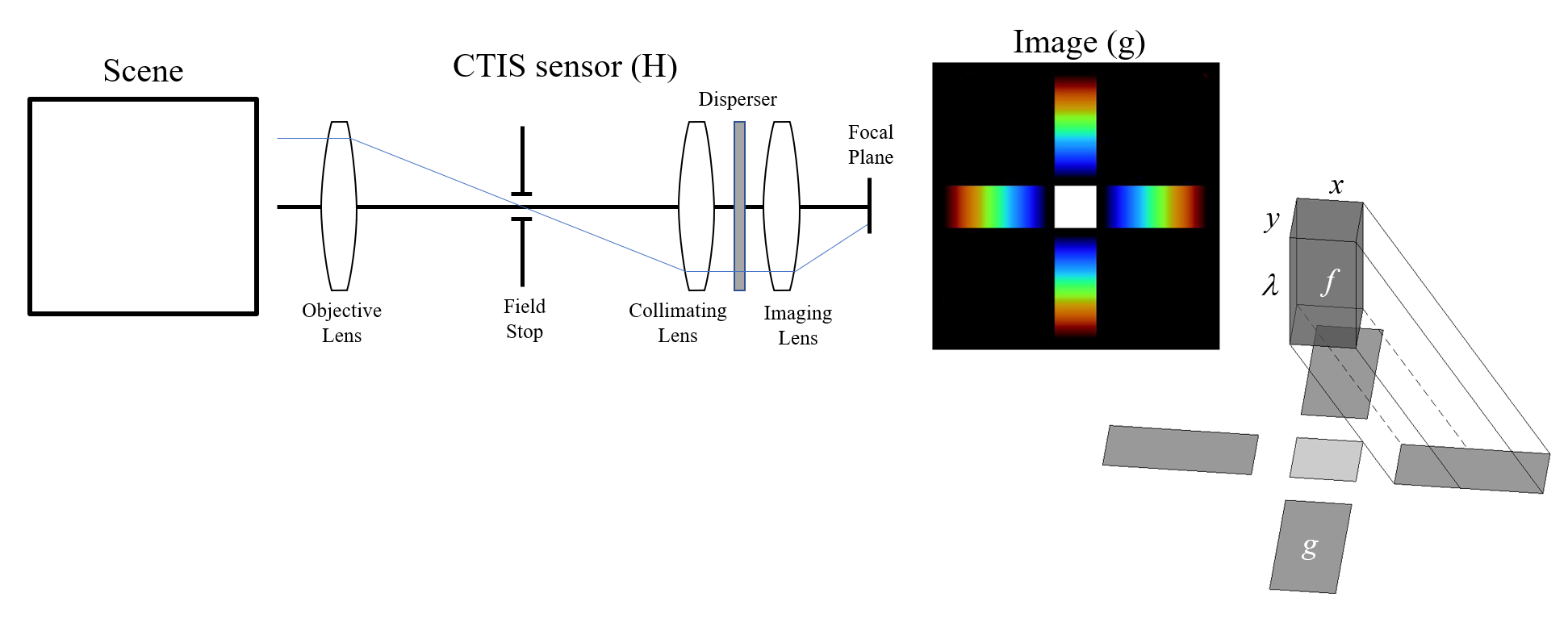}
   		\end{tabular}
   	\end{center}
   	\caption{(Color online) Diagram of a CTIS sensor along with an FPA image and reconstructed datacube.}
   	\label{fig:ctis_sensor} 
\end{figure}

\begin{figure}[H]
	\begin{center}
    	\begin{tabular}{c}
   		\includegraphics[height=7cm]{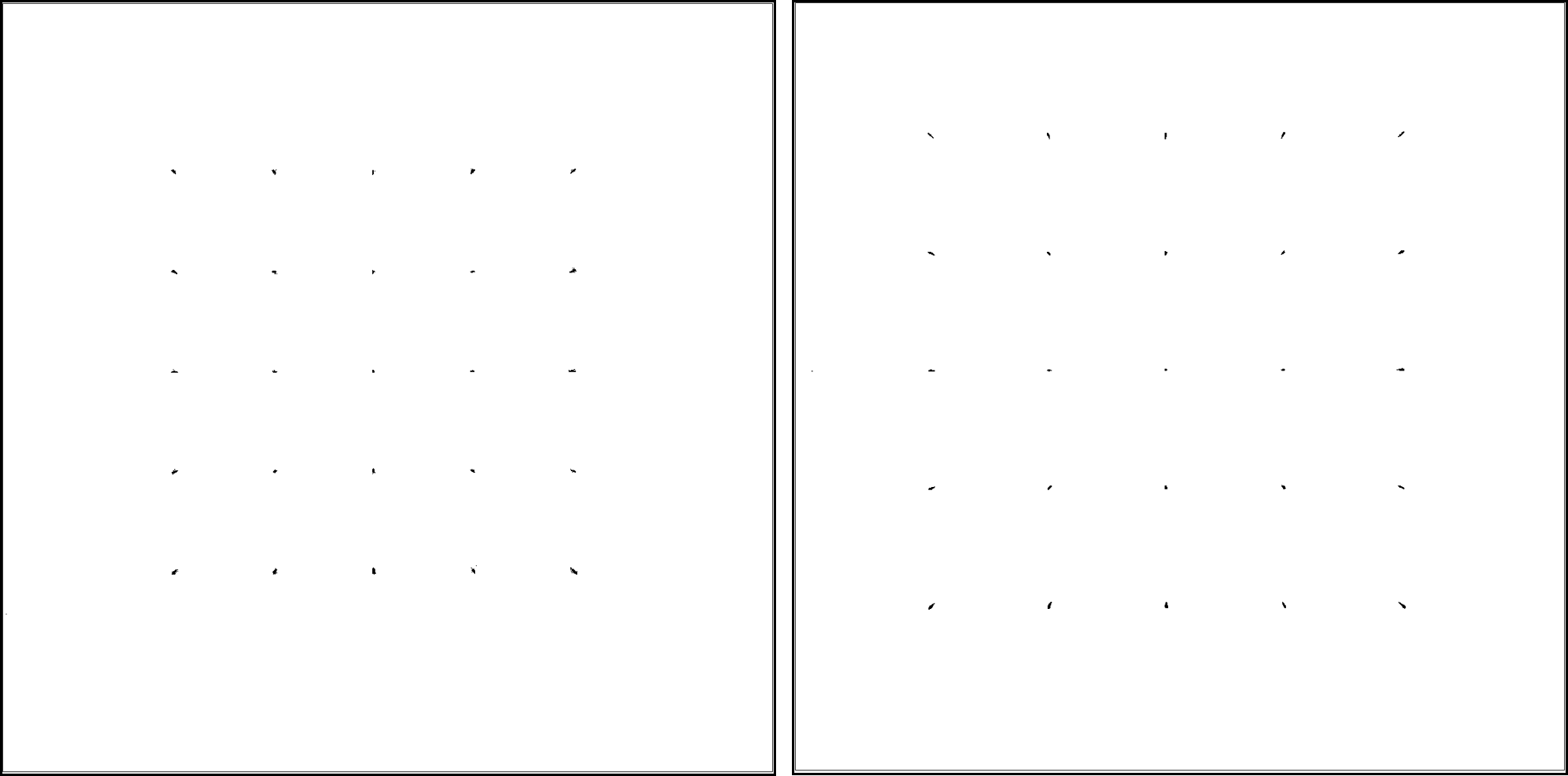}
   		\end{tabular}
   	\end{center}
   	\caption{(Color online) Two calibration images $I(x,y)$ in the upper left hand corner of the field stop. Left: $\lambda = 421$ nm and Right: $\lambda = 495$ nm.}
   	\label{fig:cal} 
\end{figure}

Fig. \ref{fig:ctis_sensor} depicts a typical CTIS snapshot hyperspectral imaging sensor $\mathbf{H}$ that encodes spatial ($x,y$) and spectral ($\lambda$) information about a scene into an image $\mathbf{g}$ on a focal plane array (FPA). The image is then computationally decoded/reconstructed to produce a hyperspectral datacube $\mathbf{f}(x,y,\lambda)$. To reconstruct the datacube $\mathbf{f}$ from an image $\mathbf{g}$ the linear imaging equation is assumed \cite{barrett2004foundations,hagen2007fourier}
\begin{equation}
	\mathbf{g} = \mathbf{H}\mathbf{f} ,
	\label{eqn:linear}
\end{equation}
where the system matrix $\mathbf{H}$ is obtained via calibration. Calibration is \hl{performed} by placing a monochromatic point source such as a small optical fiber illuminated via a monochromator at one pixel in an $a \times \alpha$ field stop. Then a set of $w$ calibration images as shown in Fig. \ref{fig:cal} are obtained on a $\gamma \times \xi$ FPA. After this, each calibration image is converted into a vector (via column major or row major order) of length $n = {\gamma}{\xi}$ and becomes one column of the system matrix. The remaining columns are obtained assuming shift-invariance where we assume for a given wavelength, shifting the optical fiber one pixel in the field stop or one voxel in the datacube will result in a calibration image that is shifted one pixel on the FPA. Doing so results in an $n \times m$ system matrix with $m = a \alpha w$. The assumption of shift-invariance is widely used and it's validity is explored by Hagen \cite{hagen2007fourier, hagen2007snapshot}.
The solution to Eq. (\ref{eqn:linear}) is formally the Moore-Penrose pseudoinverse $\mathbf{f} = \mathbf{H}^{+}\mathbf{g}$, however, since the system matrix is large and sparse, iterative solvers such as: EM/MLEM, MART, MERT, ART, etc. \hl{are often employed.} \cite{volin2000portable,shepp1982maximum,volin39,hagen2007fourier,horton2010novel,garcia1999mixed,sethaphong2007large,gordon1970algebraic} For our purposes \hl{we used} the Expectation Maximization (EM) solver
\begin{equation}
\label{eqn:em}
\mathbf{f}^{(k+1)} = \frac{\mathbf{f}^{(k)}}{\mathbf{h}} \left( \mathbf{H}^T \frac{\mathbf{g}}{\mathbf{H}\mathbf{f}^{(k)}} \right) = \left(\mathbf{f}^{(k)} \oslash \mathbf{h} \right) \odot \left( \mathbf{H}^T \left( \mathbf{g} \oslash \mathbf{H}\mathbf{f}^{(k)} \right) \right) ,
\end{equation}
where $k$ is the iteration number, $\mathbf{h}$ in component notation is $h_j = \sum_{i=1}^n \mathbf{H}_{ij} \text{ for } j = 1 \dots m$, and $\odot$ ($\oslash$) denotes the Hadamard product (division). A typical initial guess for $\mathbf{f}^{(1)}$ is either all ones \cite{sethaphong2007large,wilson1997reconstructions} or $\mathbf{H}^T \mathbf{g}$. \cite{descour1995computed} A typical total number of iterations $K$, is $10-30$  \cite{descour1995computed,sethaphong2007large,wilson1997reconstructions,hagen2007snapshot}. As Hagen pointed out, using a large number of iterations can result in increased reconstruction error and even unrealistic solutions recommending stopping the iterations early to avoid this effect.

\section{Parallel reconstruction algorithm} 
\label{sec:alg}
Our reconstruction algorithm utilizes the Vose-Horton decomposition. We start by assuming shift-invariance which means the system matrix can be partitioned as
\begin{equation}
\label{eqn:H}
\mathbf{H} = 
\left(
\begin{array}{cccc}
\mathbf{H}_1 & \mathbf{H}_2 & \dots & \mathbf{H}_w
\end{array}
\right) ,
\end{equation}
where each block $\mathbf{H}_i$ for $i = 1 \dots w$ is partitioned further into \textbf{rectangular circulant} blocks
\begin{equation}
\mathbf{H}_i =
\left(
\begin{array}{cccc}
\mathbf{T}_{i,1} & \mathbf{T}_{i,2} & \dots & \mathbf{T}_{i,\alpha}
\end{array}
\right) ,
\end{equation}
and each block $\mathbf{T}_{i,j}$ for $j = 1 \dots \alpha$ is an $n \times a$ \textbf{rectangular circulant} matrix; hence, $\mathbf{H}$ has $m = a{\alpha}w$ columns and $n = {\gamma}{\xi}$ rows. Next we define the ${\gamma} \times a$ matrix
\begin{equation}
\mathbf{Q} = 
\left(
\begin{array}{c}
	\mathbf{I}_a \\
	\mathbf{0}
\end{array}
\right) ,
\end{equation}
and the $n \times a{\alpha}$ matrix
\begin{equation}
\mathbf{E} = 
\left(
\begin{array}{c}
	\mathbf{I}_\alpha \otimes \mathbf{Q} \\
	\mathbf{0}
\end{array}
\right) ,
\end{equation}
where $\mathbf{0}$ is the zero matrix, $\mathbf{I}_p$ is the $p \times p$ identity matrix, and $\otimes$ is the Kronecker product. \cite{horton2010novel, vose2007heuristic} We now decompose the system matrix as
\begin{equation}
\label{eqn:decomposition}
\mathbf{H}_i = \mathbf{C}_i\mathbf{E} ,
\end{equation}
where $\mathbf{C}_i$ is an $n \times n$ circulant matrix whose first column $\mathbf{c}_i$ is $\mathbf{T}_{i,1}$. By inserting our results from Eq. (\ref{eqn:decomposition}) into Eq. (\ref{eqn:H}) and then into Eq. (\ref{eqn:linear}) we end up with
\begin{equation}
\label{eqn:vs_linear}
\mathbf{g} = \mathbf{H}\mathbf{f} = \sum_{i = 1}^w {\mathbf{C}_i}{\mathbf{E}}{\mathbf{f}_i} = \left( \begin{array}{cccc}
\mathbf{C}_1 & \mathbf{C}_2 & \dots & \mathbf{C}_w
\end{array} \right) \left( \mathbf{I}_w \otimes \mathbf{E} \right) \mathbf{f} ,
\end{equation}
where $\mathbf{f}$ is a block vector composed of $w$ blocks each of length $\ell = a \alpha$
\begin{equation}
\mathbf{f} = \left(
\begin{array}{c}
	\mathbf{f}_1  \\
	\vdots \\
	\mathbf{f}_w
\end{array}
\right) ;
\end{equation}
hence, $\mathbf{f}$ has length $m = \ell w = a \alpha w$.

It turns out, the matrix vector multiplication $\left( \mathbf{I}_w \otimes \mathbf{E} \right) \mathbf{f}$, \hl{is a parallel} $1-1$ index mapping. This mapping can be done \hl{quickly} on a GPU with $m$ threads. To see this, we define the vector of length $nw$ that results from the matrix vector multiplication $\mathbf{v} = \left( \mathbf{I}_w \otimes \mathbf{E} \right) \mathbf{f}$ as
\begin{equation}
\label{eqn:v}
\mathbf{v} = \left(
\begin{array}{c}
	\mathbf{v}_1  \\
	\vdots \\
	\mathbf{v}_w
\end{array}
\right) ,
\end{equation}
where each $\mathbf{v}_i = \mathbf{E}\mathbf{f}_i$ for $i = 1 \dots w$ is a vector of length $n$. Next we denote the $i^{th}$ element of $\mathbf{v}$ as $\left(\mathbf{v}\right)_i$ (and similarly for the $j^{th}$ element of $\mathbf{f}$). Then the \textbf{zero based} indexing mapping algorithm is
\begin{enumerate}
\item Initialize $\mathbf{v} = \mathbf{0}$.
\item $\left( \mathbf{v} \right)_i = \left(\mathbf{f} \right)_j \text{ for } j = 0 \dots m-1$ and $i$ is defined completely in terms of $j$ and constants as
\begin{equation}
i = j - s \ell + \left( \gamma - a \right) \left\lfloor \frac{j - s \ell}{a} \right\rfloor + sn \text{ where } s = \left\lfloor \frac{j}{\ell} \right\rfloor.
\end{equation}
\end{enumerate}
Rewriting Eq. (\ref{eqn:vs_linear}) in terms of $\mathbf{v}$, \hl{we arrive at}
\begin{equation}
\label{eqn:linear_circulant}
\mathbf{g} = \mathbf{H} \mathbf{f} = \sum_{i = 1}^w {\mathbf{C}_i}{\mathbf{v}_i} .
\end{equation}
\hl{It is} well \hl{known} that circulant matrix vector multiplication can be \hl{performed} via Fourier transforms in $\mathcal{O}(n \log n)$ time. This is because a circulant matrix $\mathbf{C}_i$ for $i = 1 \dots w$ can be written in terms of a diagonal matrix $\mathbf{D}_i$ as $\mathbf{D}_i = \mathbf{F}_1\mathbf{C}_i\mathbf{F}_1^{-1}$. In our notation, we denoted the 1D Fourier transform and \hl{its} inverse as $\mathbf{F}_1$ and $\mathbf{F}_1^{-1}$. It follows that Eq. (\ref{eqn:linear_circulant}) can be written as
\begin{equation}
\label{eqn:forward}
\mathbf{g} = \mathbf{H} \mathbf{f} = \sum_{i = 1}^w {\mathbf{F}_1^{-1}\mathbf{F}_1\mathbf{C}_i}\mathbf{F}_1^{-1}\mathbf{F}_1{\mathbf{v}_i} = \mathbf{F}_1^{-1} \sum_{i = 1}^w {\mathbf{D}_i}\mathbf{F}_1{\mathbf{v}_i} = \mathbf{F}_1^{-1} \sum_{i = 1}^w \mathbf{d}_i \odot \left( \mathbf{F}_1{\mathbf{v}_i} \right) ,
\end{equation}
where $\mathbf{d}_i = \mathbf{F}_1 \mathbf{c}_i$ is precomputed. \hl{Thus, we reduced} the expensive matrix vector multiplication $\mathbf{H}\mathbf{f}$ down to three trivial parallel operations. The \hl{processing bottleneck then becomes} the Fourier transform which can be computed using cuFFT on a GPU \cite{cudadocs}.

We now need the backward projection $\mathbf{H}^{T}\mathbf{u}$ which is found by taking the transpose of the system matrix defined by Eq. (\ref{eqn:vs_linear}) $\mathbf{H}^T =
\left( \left( \begin{array}{cccc}
\mathbf{C}_1 & \mathbf{C}_2 & \dots & \mathbf{C}_w
\end{array} \right)\left(\mathbf{I}_w \otimes \mathbf{E} \right) \right)^T = \left(\mathbf{I}_w \otimes \mathbf{E} \right)^T \left( \begin{array}{cccc}
\mathbf{C}_1 & \mathbf{C}_2 & \dots & \mathbf{C}_w
\end{array} \right)^T$. This results in
\begin{equation}
\label{eq:Htu}
\mathbf{H}^T \mathbf{u} = \left(\mathbf{I}_w \otimes \mathbf{E} \right)^T \left(
\begin{array}{c}
	\mathbf{C}_1^T \mathbf{u}  \\
	\mathbf{C}_2^T \mathbf{u} \\
	\vdots \\
	\mathbf{C}_w^T \mathbf{u}
\end{array}
\right) ,
\end{equation}
which is also composed of a trivial index mapping $\left(\mathbf{I}_w \otimes \mathbf{E} \right)^T$ that can be \hl{performed} on a GPU with $m$ threads. If we again denote the resulting vectors $\mathbf{z} =  \left( \begin{array}{cccc}
\mathbf{C}_1 & \mathbf{C}_2 & \dots & \mathbf{C}_w
\end{array} \right)^T \mathbf{u}$ and $ \boldsymbol{\zeta} = \left(\mathbf{I}_w \otimes \mathbf{E} \right)^T \mathbf{z}$ the \textbf{zero based} indexing mapping algorithm is
\begin{enumerate}
\item $\zeta_i = z_j \text{ for } i = 0 \dots m-1$ and $j$ is defined completely in terms of $i$ and constants as
\begin{equation}
j = i - s \ell + \left( \gamma - a \right) \left\lfloor \frac{i - s \ell}{a} \right\rfloor + sn \text{ where } s = \left\lfloor \frac{i}{\ell} \right\rfloor .
\end{equation}
\end{enumerate}
Since the transpose of a circulant matrix is also circulant, and because Fourier transforms along with diagonal matrices are symmetric\hl{,} we can simplify Eq. (\ref{eq:Htu}) as

\begin{equation}
\label{eqn:transpose_cir}
\mathbf{C}_i^T \mathbf{u} = \mathbf{F}_1 \left( \mathbf{F}_1 \mathbf{C}_i \mathbf{F_1}^{-1} \right)^T \mathbf{F_1}^{-1} \mathbf{u} = \mathbf{F}_1 \mathbf{D}_i \mathbf{F}_1^{-1} \mathbf{u} \text{ for } i = 1 \dots w.
\end{equation}

An additional optimization using Hermitian symmetry can also be exploited. Hermitian symmetry implies that when we compute the forward Fourier transform of a real vector of length $n$, we only need the resulting complex vector of length $\beta = \left\lfloor \frac{n}{2}\right\rfloor + 1$. Likewise, the inverse Fourier transform of a complex vector results in a real vector of length $n$, so the complex input only needs a length of $\beta$ \cite{cudadocs}. In order to leverage Hermitian symmetry we need to write Eq. (\ref{eqn:transpose_cir}) with the Fourier terms flipped. This is accomplished by noticing that, $\overline{\mathbf{F}}_1 = \mathbf{F}^{-1}_1$, $\overline{\mathbf{F}}_1^{-1} = \mathbf{F}_1$ (bar denotes complex conjugate), while $\mathbf{u}$ and $\mathbf{C}_i^T \mathbf{u}$ are real
\begin{equation}
\label{eqn:backwards}
\mathbf{C}_i^T \mathbf{u} = \overline{\mathbf{C}_i^T \mathbf{u}} = \overline{\mathbf{C}_i^T} \overline{\mathbf{u}} = \overline{\mathbf{C}_i^T} \mathbf{u} = \overline{\mathbf{F}_1 \mathbf{D}_i \mathbf{F}_1^{-1}} \mathbf{u} = \mathbf{F}_1^{-1} \overline{\mathbf{D}}_i \mathbf{F}_1 \mathbf{u} = \mathbf{F}_1^{-1} \left( \overline{\mathbf{d}}_i \odot \left( \mathbf{F}_1 \mathbf{u} \right) \right) \text{ for } i = 1 \dots w  .
\end{equation}
This substantially reduces the spacetime complexity since
\begin{enumerate}
\item $\mathbf{d}_i = \mathbf{F}_1 \mathbf{c}_i$ with $\mathbf{c}_i$ real implies we only need the first $\beta$ elements of $\mathbf{d}_i$ instead of $n$ elements.

\item $\mathbf{F}_1 \mathbf{v}$ ($\mathbf{F}_1 \mathbf{u}$) result in vectors of length $\beta$ which means the Hadamard product between them and $\mathbf{d}_i$ ($\overline{\mathbf{d}}_i$) is \hl{performed} on $\beta$ elements instead of $n$ elements.
\end{enumerate}

Just like the forward projection, we \hl{reduced} the expensive matrix vector multiplication $\mathbf{H}^T\mathbf{u}$ down to two simple parallel operations leaving the bottleneck as the Fourier transform since the remaining Hadamard operations $\mathbf{g} \oslash (\dots)$ and $\left(\mathbf{f} \oslash \mathbf{h} \right) \odot \left( \dots \right)$ are trivially parallelizable. 

For convenience we provided pseduocode for our parallel algorithm (hereafter referred to as WBH) implemented for the EM solver in Alg. \ref{alg:wbh}. Note that, even though we applied our algorithm to the EM solver, the WBH algorithm is solver agnostic. Finally, the complexity analysis of the WBH algorithm is driven by the Fourier transform. Since cuFFT \cite{cudadocs} has a guaranteed runtime of $\mathcal{O}(n \log n)$, we conclude our complexity is $\mathcal{O}\left( knw \log n \right)$. This is in contrast to the $\mathcal{O}\left( knw \ell \right)$ runtime using a brute-force (hereafter referred to as BF) matrix vector multiplication algorithm. In addition, the space complexity of WBH is $\mathcal{O}\left( n w \right)$ which is also better than the $\mathcal{O}\left( nw \ell \right)$ BF method. 

\begin{algorithm}[H]
\label{alg:wbh}
\DontPrintSemicolon
\caption{WBH algorithm implemented in the EM solver. Each line in the for loop is a kernel call.}

Read and copy $\mathbf{d_1}, \mathbf{d_2}, \dots \mathbf{d}_w$ to the GPU \;
Read and copy $\mathbf{g}$ to the GPU \;
$K =$ some value between $10-30$ \;
Initialize $\mathbf{f}^{(1)}$ to all ones \;
\For{ k = 1 \KwTo K}{
	$\mathbf{v} = \left( \mathbf{I}_w \otimes \mathbf{E} \right) \mathbf{f}^{(k)}$ \;
	$\mathbf{g}^{(k)} = \mathbf{F}_1^{-1} \left( \mathbf{d}_1 \odot 	\left( \mathbf{F}_1 \mathbf{v}_1 \right) + \mathbf{d}_2 \odot 	\left( \mathbf{F}_1 \mathbf{v}_2 \right) + \dots + \mathbf{d}_w \odot 	\left( \mathbf{F}_1 \mathbf{v}_w \right) \right)$ \;
	$\mathbf{u} = \mathbf{g} \oslash \mathbf{g}^{(k)}$ \;
	$\mathbf{U} = \mathbf{F}_1 \mathbf{u}$	 \;
	$ \mathbf{z} = \mathbf{F}_1^{-1} \left( \begin{array}{cccc}\overline{\mathbf{d}}_1 \odot \mathbf{U} & \overline{\mathbf{d}}_2 \odot \mathbf{U} & \dots & \overline{\mathbf{d}}_w \odot \mathbf{U} \end{array} \right)^T$ \;
	 $\boldsymbol{\zeta} = \left( \mathbf{I}_w \otimes \mathbf{E} \right)^T \mathbf{z}$ \;
	 $\mathbf{f}^{(k+1)} = \left(\mathbf{f}^{(k)} \odot \boldsymbol{\zeta} \right) \oslash \mathbf{h}$ \;
}

Copy $\mathbf{f}^{(k+1)}$ from the GPU to the CPU \;
\textbf{Save:} $\mathbf{f}^{(k+1)}$ \;

\end{algorithm}

\section{Results and Discussion}
\begin{figure}[H]
	\begin{center}
    	\begin{tabular}{c}
   		\includegraphics[height=7cm]{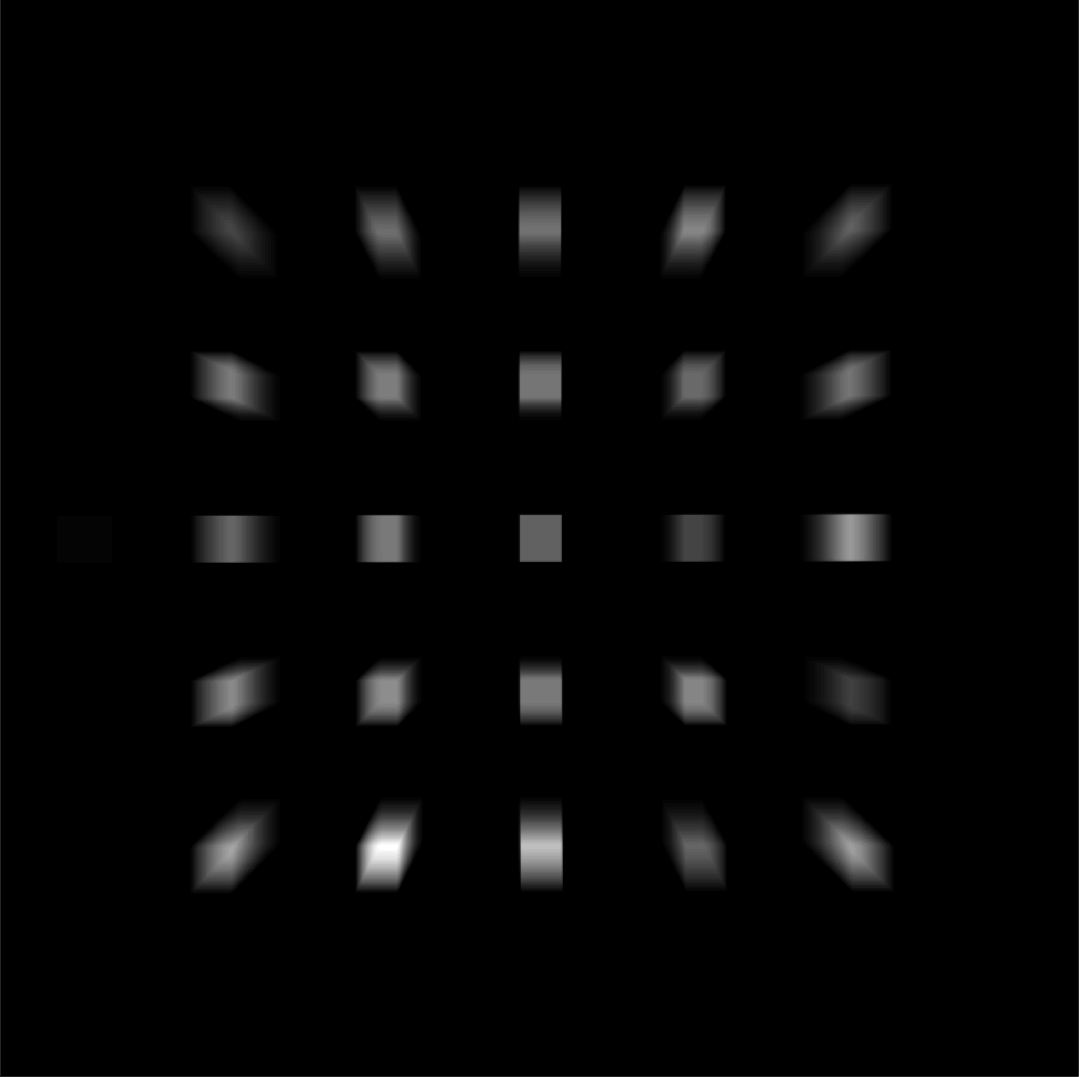}
   		\end{tabular}
   	\end{center}
   	\caption{(Color online) FPA image generated using a fully illuminated field stop $\mathbf{f}(x,y,\lambda) = 100$.}
   	\label{fig:g} 
\end{figure}

Using a system matrix provided by Sethaphong which spans $421$ nm up to $495$ nm in steps of $1$ nm we synthetically generated an image using a fully illuminated ($\mathbf{f}(x,y,\lambda) = 100$) $89 \times 80$ field stop ($a = 89$ and $\alpha = 80$) on a $2048 \times 2048$ FPA ($\gamma = \xi = 2048$) \hl{. The FPA image is shown in Fig. {\ref{fig:g}}}.

Our benchmarks in seconds are shown in Table \ref{tbl:bm} and were obtained using the following hardware:
\begin{enumerate}
\item Desktop computer with a modest Quadro P4000 GPU and $8$ GB of GPU memory. We implemented the WBH + EM + P4000 (WBH implementation of the EM algorithm and running on a P4000 GPU) algorithm using CUDA C/C++ $10.1$.
\item Nvidia Jetson AGX Xavier (hereafter referred to as Xavier) with an embedded $512$-core Volta GPU and 16 GB of GPU memory \cite{jetson}. \hl{We will} refer to this benchmark as WBH + EM + Xavier (WBH implementation of the EM algorithm running on an Xavier GPU).
\item Desktop computer with \textbf{two} 2.30 GHz Intel(R) Xeon(R) E5-2699 v3 CPUs and $\mathbf{256}$ GB of RAM. We implemented the BF + EM + CPU (BF implementation of the EM algorithm running on a CPU) algorithm using Matlab.
\end{enumerate}

\begin{table}[H]
\centering
\begin{tabular}{rrrrr}
\hline
\multicolumn{1}{r}{\bfseries BF + EM + CPU (s)} & \multicolumn{1}{r}{\bfseries WBH + EM + Xavier (s)} & \multicolumn{1}{c}{\bfseries WBH + EM + P4000 (s)} & \multicolumn{1}{r}{\bfseries $K$} & \multicolumn{1}{r}{\bfseries $w$} \\
\hline
21593.7 & 853.0 & 467.2 & 3100 & 75 \\
438.5 & 13.9 &  7.5 &   50 & . \\
315.7 & 9.2 &  5.0 &   33 & . \\
265.3 & 7.0 &  3.8 &   25 & . \\
161.0 & 2.8 &  1.5 &   10 & . \\

\hline
7537.2 & 281.1 & 153.2 & 3100 & 24 \\
149.1 & 4.6 &  2.5 &   50 & . \\
109.7 & 3.0 &  1.6 &   33 & . \\
91.9 & 2.3 &  1.2 &   25 & . \\
53.6 & 0.9 &  0.5 &   10 & . \\
\hline
1039.3 & 41.3 & 24.0 & 3100 & 3 \\
20.2 & 0.7 & 0.4 &   50 & . \\
13.9 & 0.5 & 0.3 &   33 & . \\
11.8 & 0.4 & 0.2 &   25 & . \\
6.8 & 0.1 & 0.08 &  10 & . \\

\hline
\end{tabular}
\caption{Runtime comparisons in seconds for BF + EM + CPU, WBH + EM + Xavier, WBH + EM + P4000 with a given total number of iterations $K$ and total number of wavelengths $w$.}
\label{tbl:bm}
\end{table}

As can be seen from Table. \ref{tbl:bm} the WBH algorithm \hl{performs} well. For example, in the case $w = 75, K = 25$ the runtime is $3.8$ s using a desktop GPU and $7.0$ s using an embedded GPU. Furthermore, the relative error was $0.02$. In comparison, the BF algorithm took $265$ s and had a relative error of $1.4 \times 10^{-16}$. One explanation for the higher error is the use of single precision to save space when running WBH as opposed to double precision when running the BF method.

We can also compare our results to Vose-Horton. They used a CTIS system similar to ours but with $24$ wavelengths and ran their novel heuristic reconstruction algorithm on a CPU. Their reconstruction quality/error was excellent and it took $423.0$ s. They mentioned that choosing $K = 33$ for the BF MART solver did a fair job at reconstructing the image but the solver needed to be run for $K=3100$ in order to get as good of results as them. Notice that, in both cases $K = 33$ and $K=3100$, our algorithm ran faster. Furthermore, as discussed in Sec. \ref{sec:ctisoverview}, $10-30$ is a typical total number of iterations. For these reasons, we conclude the WBH algorithm when applied to the EM solver runs in a few seconds.

In addition to Vose-Horton, Sethaphong, using the same CTIS system as us ran a BF parallel EM reconstruction algorithm on a \hl{dedicated} set of 10 processors that was part of a larger 1500 core supercomputer. For his benchmarks he used $w = 280$, $K=50$ and obtained a time of $37.08$ s. Unfortunately, due to memory limitations, we only \hl{processed} $75$ wavelengths. However, we have table entries at $(w = 24, K = 50)$, $(w = 75,K = 50)$ and our runtime is linear in the total number of wavelengths. If we \hl{linearly extrapolate} for $w=280$ we \hl{obtain} $27.6$ s. This is, theoretically, faster than Sethaphong.

\begin{figure}[H]
	\begin{center}
    	\begin{tabular}{c}
   		\includegraphics[height=7cm]{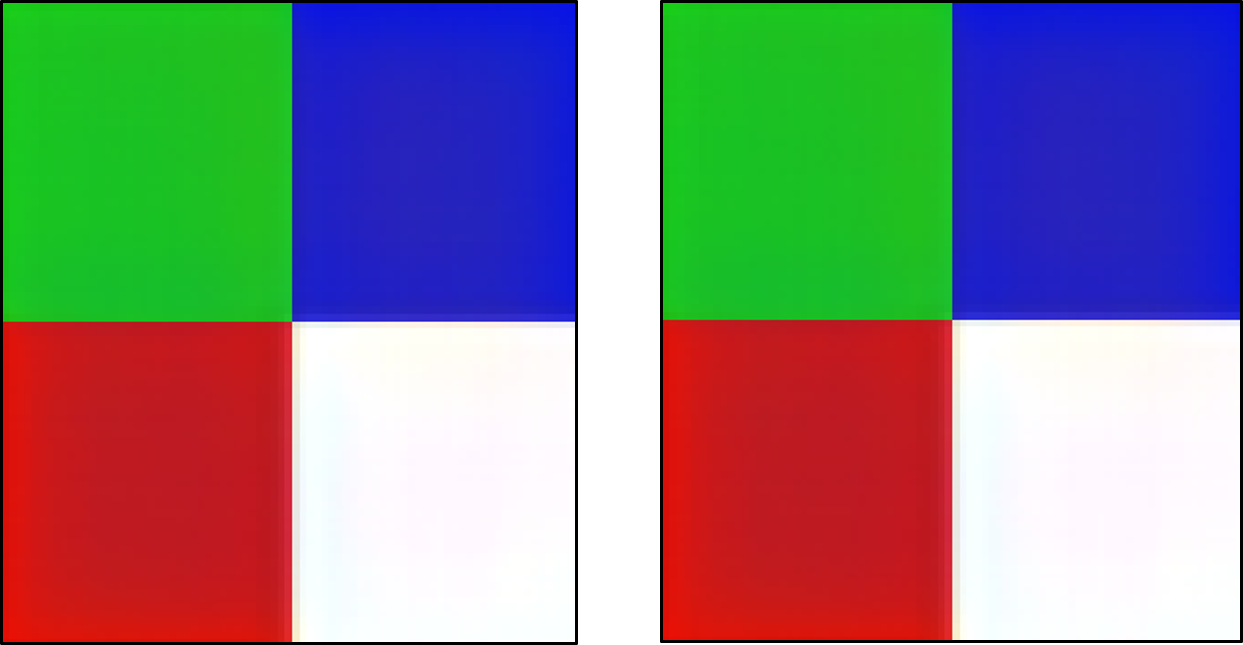}
   		\end{tabular}
   	\end{center}
   	\caption{\hl{(Color online) RBG images used to confirm our reconstruction quality. The images corresponded to a datacube with three wavelengths and where reconstructed using $33$ iterations. Left: Reconstructed datacube using the EM method. Right: Reconstructed datacube using the WBH algorithm.}\label{fig:recon}}
\end{figure}

\hl{
We validated our reconstruction quality using a method similar to Vose-Horton. Using an RGB image that corresponds to a datacube with three wavelengths, we multiplied the vectorized RBG image by our system matrix which produced an FPA image that was fed into both the EM and WBH algorithms. The reconstructed results using $33$ iterations in the aforementioned algorithms are shown in the left and right of Fig. {\ref{fig:recon}} respectively. As can be seen from Fig. {\ref{fig:recon}}, the reconstructed datacube using the WBH algorithm looks the same as the one using the EM method. Quantitatively, the average relative pixel error defined as $\frac{1}{m}\sum_{j=1}^m \frac{|f^{wbh}_j - f^{em}_j|}{f^{em}_j}$ where $\mathbf{f}^{em}$($\mathbf{f}^{wbh}$) denotes the vectorized left(right) RGB image in Fig. {\ref{fig:recon}} was found to be $0.5 \times 10^{-3}$. Thus, we obtained excellent reconstruction quality.
}

It would be interesting as Horton suggested in his thesis \cite{horton2010novel} to implement the Vose-Horton algorithm on a GPU. One \hl{benefit besides a potential processing gain}, is the possibility to use a hybrid approach that utilizes both algorithms. This is feasible because the WBH and Vose-Horton algorithm share the same matrix decomposition. In fact, \hl{Thompson implemented} hardware acceleration for the Vose-Horton algorithm on a cell processor (running on \hl{a} Playstation 3). Unfortunately, due to memory limitations he was only able to \hl{process} three wavelengths but, the results are still interesting. Thompson showed, with hardware acceleration the Vose-Horton algorithm could run in $4.14$ s. In comparison, the WBH algorithm for $w = 3, K=3100$ was slower. However, in the more common case $w = 3, K=33$ the WBH algorithm was faster. Given that cell processors are less popular than GPUs today, we believe a GPU implementation is a better approach. Finally, Thompson's issues with memory bring up a good point. For embedded systems and GPUs, memory is a commodity. The fact that we were able to run $75$ wavelengths on a modest $8$ GB GPU is an excellent result. If we were to use a high end desktop with better and more GPUs, we think \hl{it is} possible to run as many wavelengths as Sethaphong while still maintaining a few second reconstruction time. Overall, we developed a simple, fast, and solver agnostic algorithm \hl{that is} practical for near-realtime and/or embedded systems.

\section{Conclusions}
We have developed a parallel algorithm that accelerated CTIS reconstruction times and exploited spatial shift invariance. The new algorithm, WBH, was applied to the EM solver and implemented \hl{on a desktop and embedded GPU. After validating that our new algorithm maintained reconstruction quality on par with traditional methods we went on to benchmark the new WBH algorithm which was shown to preform well; achieving reconstruction times of a few seconds where traditional brute-force methods required minutes to hours.} We also demonstrated the feasibility of GPUs for CTIS reconstruction.

\bibliography{bibfile}

\begin{thebibliography}{10}

\bibitem{horton2010novel}
Horton, M.~D., ``A novel technique for ctis image-reconstruction,'' (2010).

\bibitem{descour1995computed}
Descour, M. and Dereniak, E., ``Computed-tomography imaging spectrometer:
  experimental calibration and reconstruction results,'' {\em Applied
  optics}~{\bf 34}(22),  4817--4826 (1995).

\bibitem{vane1988terrestrial}
Vane, G. and Goetz, A.~F., ``Terrestrial imaging spectroscopy,'' {\em Remote
  Sensing of Environment}~{\bf 24}(1),  1--29 (1988).

\bibitem{sabatke2002snapshot}
Sabatke, D.~S., Locke, A.~M., Dereniak, E.~L., Descour, M.~R., Garcia, J.~P.,
  Hamilton, T.~K., and McMillan, R.~W., ``Snapshot imaging
  spectropolarimeter,'' {\em Optical engineering}~{\bf 41} (2002).

\bibitem{Azzam1995}
Azzam, R. M.~A.,  [{\em Ellipsometry, Handbook of
  Optics}{\nolinebreak\hspace{0.1em}]}, McGraw Hill, 2 (m. bass, ed.)~ed.
  (1995).

\bibitem{chipman1995}
Chipman, R.~A.,  [{\em Polarimetry, Handbook of
  Optics}{\nolinebreak\hspace{0.1em}]}, McGraw Hill, 2 (m. bass, ed.)~ed.
  (1995).

\bibitem{Henderson1995}
Henderson, B.,  [{\em Optical spectrometers, Handbook of
  Optics}{\nolinebreak\hspace{0.1em}]}, McGraw Hill, 2 (m. bass, ed.)~ed.
  (1995).

\bibitem{Horton24}
 [{\em Spectoscopic Measurements, Handbook of
  Optics}{\nolinebreak\hspace{0.1em}]}, McGraw Hill, 2 (m. bass, ed.)~ed.
  (1995).

\bibitem{scholl2003phase}
Scholl, J.~F., Dereniak, E.~L., Descour, M.~R., Tebow, C.~P., and Volin, C.~E.,
  ``Phase grating design for a dual-band snapshot imaging spectrometer,'' {\em
  Applied optics}~{\bf 42}(1),  18--29 (2003).

\bibitem{luo2007fast}
Luo, Y., Liao, N., Wang, X., Liang, M., and Fen, J., ``Fast processing of
  imaging spectrometer data cube based on fpga design,'' in [{\em MIPPR 2007:
  Multispectral Image Processing}{\nolinebreak\hspace{0.1em}]},   {\bf 6787},
  678708, International Society for Optics and Photonics (2007).

\bibitem{descour1997demonstration}
Descour, M.~R., Volin, C.~E., Dereniak, E.~L., Gleeson, T.~M., Hopkins, M.~F.,
  Wilson, D.~W., and Maker, P.~D., ``Demonstration of a computed-tomography
  imaging spectrometer using a computer-generated hologram disperser,'' {\em
  Applied Optics}~{\bf 36}(16),  3694--3698 (1997).

\bibitem{goetz1985imaging}
Goetz, A.~F., Vane, G., Solomon, J.~E., and Rock, B.~N., ``Imaging spectrometry
  for earth remote sensing,'' {\em science}~{\bf 228}(4704),  1147--1153
  (1985).

\bibitem{johnson2007snapshot}
Johnson, W.~R., Wilson, D.~W., Fink, W., Humayun, M.~S., and Bearman, G.~H.,
  ``Snapshot hyperspectral imaging in ophthalmology,'' {\em Journal of
  biomedical optics}~{\bf 12}(1),  014036 (2007).

\bibitem{vo2014biomedical}
Vo-Dinh, T.,  [{\em Biomedical Photonics
  Handbook}{\nolinebreak\hspace{0.1em}]}, CRC press (2002).

\bibitem{murguia2000compact}
Murguia, J.~E., Reeves, T.~D., Mooney, J.~M., Ewing, W.~S., Shepherd, F.~D.,
  and Brodzik, A.~K., ``Compact visible/near-infrared hyperspectral imager,''
  in [{\em Infrared Detectors and Focal Plane Arrays
  VI}{\nolinebreak\hspace{0.1em}]},   {\bf 4028},  457--468, International
  Society for Optics and Photonics (2000).

\bibitem{wilson2005reconstruction}
Wilson, M.~P., Ford, B.~K., and Salazar, J. S.~I., ``Reconstruction algorithm
  development and assessment for a computed tomography based-spectral
  imager.,'' tech. rep., Sandia National Laboratories (2005).

\bibitem{scholl2006evaluations}
Scholl, J.~F., Hege, E.~K., Lloyd-Hart, M., O'Connell, D., Johnson, W.~R., and
  Dereniak, E.~L., ``Evaluations of classification and spectral unmixing
  algorithms using ground based satellite imaging,'' in [{\em Algorithms and
  Technologies for Multispectral, Hyperspectral, and Ultraspectral Imagery
  XII}{\nolinebreak\hspace{0.1em}]},   {\bf 6233},  623328, International
  Society for Optics and Photonics (2006).

\bibitem{hagen2007fourier}
Hagen, N., Dereniak, E.~L., and Sass, D.~T., ``Fourier methods of improving
  reconstruction speed for ctis imaging spectrometers,'' in [{\em Imaging
  Spectrometry XII}{\nolinebreak\hspace{0.1em}]},   {\bf 6661},  666103,
  International Society for Optics and Photonics (2007).

\bibitem{hege2004hyperspectral}
Hege, E.~K., O'Connell, D., Johnson, W., Basty, S., and Dereniak, E.~L.,
  ``Hyperspectral imaging for astronomy and space surveillance,'' in [{\em
  Imaging Spectrometry IX}{\nolinebreak\hspace{0.1em}]},   {\bf 5159},
  380--391, International Society for Optics and Photonics (2004).

\bibitem{descour1994non}
Descour, M.~R., ``Non-scanning imaging spectrometry.,'' (1994).

\bibitem{volin1998high}
Volin, C.~E., Ford, B.~K., Descour, M.~R., Garcia, J.~P., Wilson, D.~W., Maker,
  P.~D., and Bearman, G.~H., ``High-speed spectral imager for imaging transient
  fluorescence phenomena,'' {\em Applied optics}~{\bf 37}(34),  8112--8119
  (1998).

\bibitem{thompson2009accelerated}
Thompson, T.~J., ``Accelerated ctis using the cell processor,'' {\em Masters
  Theses} ,  564 (2009).

\bibitem{vose2007heuristic}
Vose, M.~D. and Horton, M.~D., ``A heuristic technique for ctis image
  reconstruction,'' {\em Applied optics}~{\bf 46}(26),  6498--6503 (2007).

\bibitem{sethaphong2007large}
Sethaphong, L., ``Large format ctis in real time: Parallelized algorithms and
  preconditioning initializers,'' (2007).

\bibitem{barrett2004foundations}
Barrett, H.~H. and Myers, K.,  [{\em Foundations of Image
  Science}{\nolinebreak\hspace{0.1em}]}, John Wiley \& Sons (2004).

\bibitem{hagen2007snapshot}
Hagen, N., ``Snapshot imaging spectropolarimetry,'' (2007).

\bibitem{volin2000portable}
Volin, C.~E., ``Portable snapshot infrared imaging spectrometer,'' (2000).

\bibitem{shepp1982maximum}
Shepp, L.~A. and Vardi, Y., ``Maximum likelihood reconstruction for emission
  tomography,'' {\em IEEE transactions on medical imaging}~{\bf 1}(2),
  113--122 (1982).

\bibitem{volin39}
Lent, A., ``A convergent algorithm for maximum entropy image restoration,''
  {\em SPSE Conference Proceedings} ,  249--257 (July 1976).

\bibitem{garcia1999mixed}
Garcia, J.~P. and Dereniak, E.~L., ``Mixed-expectation image-reconstruction
  technique,'' {\em Applied optics}~{\bf 38}(17),  3745--3748 (1999).

\bibitem{gordon1970algebraic}
Gordon, R., Bender, R., and Herman, G.~T., ``Algebraic reconstruction
  techniques (art) for three-dimensional electron microscopy and x-ray
  photography,'' {\em Journal of theoretical Biology}~{\bf 29}(3),  471--481
  (1970).

\bibitem{wilson1997reconstructions}
Wilson, D.~W., Maker, P.~D., and Muller, R.~E., ``Reconstructions of
  computed-tomography imaging spectrometer image cubes using calculated system
  matrices,'' in [{\em Imaging Spectrometry III}{\nolinebreak\hspace{0.1em}]},
   {\bf 3118},  184--193, International Society for Optics and Photonics
  (1997).

\bibitem{cudadocs}
Nvidia, CUDA Toolkit Documentation v10.1.243,
  https://docs.nvidia.com/cuda/archive/10.1/.

\bibitem{jetson}
Nvidia, Jetson AGX Xavier,
  www.nvidia.com/en-us/autonomous-machines/embedded-systems/jetson-agx-xavier/.

\end{thebibliography}
\bibliographystyle{spiebib}

\section*{BIOGRAPHIES}
\begin{description}
\item[\hl{Larz White}] \hl{received his BS in mathematics/physics from Western Washington University and his PhD in physics from the University of Idaho. He has been with Lockheed Martin Advanced Development Programs in Fort Worth since 2017 developing image and radar algorithms. He is also a professor of computer science and engineering at the University of Texas at Arlington.}

\item[\hl{W. Bryan Bell}] \hl{received his BSEE and MSEE from Texas A{\&}M University and PhD in EE from the University of Texas at Arlington. He is currently a Senior Fellow with Lockheed Martin Advanced Development Programs in Fort Worth and a professor with TCU and UTEP. He is currently developing algorithms in targeting, geodesy, navigation fusion, combat ID, multi-sensor fusion, data exploitation, image fusion and image system development for tactical fighter aircraft.}

\item[\hl{Ryan Haygood}] \hl{received his BS in astrophysics from Texas Christian University. He has been with Lockheed Martin Aeronautics in Fort Worth, TX since 2005 performing optical design, testing, and analysis.}
\end{description}

\end{document}